\documentclass[12pt]{article}
\usepackage{amssymb}
\usepackage{amscd}

\begin{document}
\newcommand {\be}{\begin{equation}}
\newcommand {\ee}{\end{equation}}
\newcommand {\ba}{\begin{eqnarray}}
\newcommand {\ea}{\end{eqnarray}}
\newcommand {\bea}{\begin{array}}
\newcommand {\cl}{\centerline}
\newcommand {\eea}{\end{array}}
\renewcommand {\theequation}{\thesection.\arabic{equation}}
\renewcommand {\thefootnote}{\fnsymbol{footnote}}
\newcommand {\newsection}{\setcounter{equation}{0}\section}
\def \parth {{\partial\over\partial\theta_{\mu\nu}}}
\def \parmu {\partial_{\mu}}
\def \parnu {\partial_{\nu}}
\def \ola  {\overleftarrow}
\def \ora  {\overrightarrow}
\def \zz  {{\mathbb Z}}
\def \rr  {{\mathbb R}}

\def \com {commutative }
\def \ncy {noncommutativity }
\def \CS {Chern-Simons }
\def \nc {noncommutative }

\thispagestyle{empty}
\setcounter{page}0
 
\baselineskip 0.65 cm
\begin{flushright}
IC/2001/5 \\
hep-th/0102092
\end{flushright}
\begin{center}
{\Large{\bf A Note on Noncommutative Chern-Simons Theories}}
\vskip 1cm

{\bf {M.M. Sheikh-Jabbari}}

\vskip .5cm
    
{\it The Abdus Salam International Center for Theoretical Physics \\
Strada Costiera 11, Trieste, Italy}\\
{\tt jabbari@ictp.trieste.it}

\end{center}

\vskip 2cm
\begin{abstract}
The three dimensional \CS theory on  $\rr^2_{\theta}\times \rr$ is
studied. Considering the gauge transformations under the group elements
which are going to one at infinity, we show that under arbitrary
(finite) gauge transformations action changes with an integer multiple of
$2\pi$ {\it if}, the level of \nc \CS is {\it quantized}.
We also  briefly discuss the case of the  \nc torus and some other possible
extensions.

\end{abstract}
\newpage
\newsection {Introduction}

The \CS theories have been considered in mathematics and physics literature extensively. From
the physics side they shed light on the planar physics -physics in two spatial directions-
and in particular quantum Hall effect and superconductivity, for a review see
e.g. \cite{{Chern},{QHE}}.
From  the mathematical point of view, being a topological theory, it has been used to give a
"physical" interpretation to topological invariants of knot theory and
Jones polynomials \cite{{Witt},{W-D}}.

The pure \CS theory is described by the action:
\be\label{CSact}
S_{\rm {CS}}={1\over 4\pi\nu}\ \int_M \epsilon^{\mu\nu\alpha} {\rm Tr}\left(
A_{\mu}\parnu A_{\alpha} +{2i\over 3} A_{\mu} A_{\nu} A_{\alpha}\right)\ ,
\ee
where $M$ is (an oriented) 3-manifold and $A_{\mu}$ are the connections
corresponding to
a simple compact gauge group $G$.
As we can see from the action the metric of the manifold $M$ is not appearing
in (\ref{CSact}). The factor ${1\over \nu}$ (the \CS coupling) is usually called the
"level" of the \CS theory and the gauge invariance of the action under finite gauge
transformations implies that is should be
quantized, i.e. ${1\over\nu}\in \zz$ \cite{Jackiw} which is a reflection of a general 
mathematical fact that the group of continuous maps $M\to G$ is not connected. In the
homotopical classification of such maps one meets the fact that
$\Pi_3(G)=\zz$ for any simple compact group $G$.

As a field theory one can study action (\ref{CSact}). The classical equations of
motion are $F_{\mu\nu}=\partial_{[\nu}A_{\mu]}-i A_{[\mu}A_{\nu]}=0$,
the classical paths are flat connections.  In fact it
turns out that the partition
function of this theory can be expressed in terms of moduli parameters of the flat
connections \cite{Witt}.

The Yang-Mills-\CS theory is obtained by adding the usual Yang-Mills
action to 
$S_{\rm CS}$:
$$
S={4\pi\over g^2}\ \int_M  {\rm Tr}\left(F_{\mu\nu}F^{\mu\nu}\right)+
{1\over 4\pi\nu}\ \int_M \epsilon^{\mu\nu\alpha} {\rm Tr}\left(
A_{\mu}\parnu A_{\alpha} +{2i\over 3} A_{\mu} A_{\nu} A_{\alpha}\right)\ ,
$$
where $F_{\mu\nu}$ is the field strength. The pure \CS theory is then recovered at the
strong coupling limit of the Yang-Mills part ($g\to \infty$) while keeping the $\nu$ (or
the \CS coupling) fixed. The \CS term in the Yang-Mills-\CS theory can be interpreted as a
(topological) mass term for the Yang-Mills part with effective mass $m={1\over (2\pi
g)^2\nu}$ \cite{Jackiw}.
However, in order to perform the usual field theory calculations one
needs to fix the gauge symmetry. 
Although the gauge fixing terms do depend on the metric chosen on $M$, the full
quantum theory (the partition function) remains topological
\cite{{Witt},{B-T1},{B-T2}}. Perturbative loop calculations have been
performed
for both pure \CS \cite{Martin} and for Yang-Mills-\CS \cite{Rao}. Their results indicate
that the \CS level for $SU(N)$ gauge theory is renormalized as:
\be\label{levelren} 
{1\over \nu_{{\rm ren}}}={1\over \nu_{{\rm class}}}+ N\ .
\ee
Moreover it has been shown that this one loop result does not receive any
further corrections
from the higher loops \cite{{Chen},{Das}}. In fact this exact result
has been proved using the BRST (ward identity) and also the vector
supersymmetry of \CS theory \cite{Das}. This is
remarkable because it assures the level quantization to all loops order.

The field theories on the \nc spaces (in particular \nc Moyal plane and
\nc tori) has been considered a few years ago, e.g. see
\cite{Jerzy}. However, before a new re-motivation from string theory
\cite{SW}, it was not studied extensively.
Generally the \nc version of a given field theory is obtained by
replacing the usual product of the functions (fields) by the Moyal $\star$-product,
namely
\ba\label{starp}
f(x)\star g(x)&=&{\rm exp}({i\over
2}\theta^{\mu\nu}{\partial\over \partial\xi^{\mu}}{\partial\over
\partial\zeta^{\nu}})\ f(x+\xi)g(x+\zeta)\mid_{\xi=\zeta=0}\cr 
&=&
f(x)\ {\rm exp}({i\over 2}\theta^{\mu\nu}\ola{\parmu}\;\ora{\parnu})g(x)\ ,
\ea
where $\theta^{\mu\nu}$ is a constant anti-symmetric tensor.
The Moyal bracket is then defined as
\be\label{MB}
\{f,g\}=f\star g-g\star f\ .
\ee
It is easily seen that $\{x^{\mu},x^{\nu}\}=i\theta^{\mu\nu}$. 
For a review on \nc field theories and some helpful identities on $\star$-product 
see \cite{Andrei}. In particular one can construct \nc version of the
$U(N)$ gauge theories, which give rise to the \nc deformed version of
the usual gauge symmetry
\cite{SW}. We should remember that so far only the \nc $U(N)$ theory have
been considered perturbatively in the existing literature.

The \nc extension of the \CS (NCCS) theories have also been considered
\cite{{Thomas},{Chu},{Bichl},{Wu},{Poly},{Solit}}. Physically the \nc
($U(1)$) \CS theory seems to serve
a natural description for the Wigner crystal-quantum Hall fluid phase transition
\cite{Susskind}. In the context of quantum Hall effect, the parameter $\nu$
(inverse of the "\CS level") appears to be the {\it filling fraction} for quantum
Hall states. Performing the quantization in the matrix (discrete) description of 
\nc ($U(1)$) \CS, it has been shown that the factor $\nu$ is basically related to
the statistics of the particles described by NCCS action.
More precisely upon exchange of any two particles in the corresponding $n$ particle
state, we obtain a factor of ${\rm exp}({i\pi\over \nu})$:
\be\label{anyon}
\Psi(1,2,...,i,...,j,...,n)={\rm exp}\left({{i\pi\over\nu}}\right)\
\Psi(1,2,...,j,...,i,...,n)\ ,
\ee
where $\Psi(1,2,..,i,...,j,...,n)$ is an $n$ particle state. Therefore the 
NCCS particles are in general anyonic, if $\nu^{-1}$ is an arbitrary real
number \cite{Susskind}. We note that all the \nc \CS
theories with their coupling ${1\over\nu}$ differing by an {\it even}
integer, describe the same statistics. 

In this work, we study the behaviour of the ($U(1)$ and $U(N)$) \CS
theories on $\rr^2_{\theta}\times\rr$,  $\rr^2_{\theta}$ being the \nc
Moyal plane, under both infinitesimal and finite gauge transformations. We
show that the partition function (quantum theory) is invariant
under the gauge transformations, provided that ${1\over \nu}$ is an
integer. In the usual \CS terminology this means that we again face the level 
quantization. From the \nc \CS particles point of view, this means that 
NCCS describes only fermions or bosons, and not anyons.

The paper is organized as follows. In section 2, we briefly review the results of
\cite{{Bichl},{Wu}} on the perturbative analysis of NCCS theories. In
section 3, which contains our main result, we study the invariance of NCCS
action for arbitrary
(finite) gauge transformations and show that the full quantum action is
invariant, if the level is quantized.
The last section is devoted to discussions and open questions. 

\newsection{Noncommutative Chern-Simons theories: \\[-0.6cm] 
perturbative analysis}

In this section we review the results of loop calculations for \nc \CS
theories \cite{{Bichl},{Wu}}. 
The action for the \nc extension of \CS theory is
\be\label{NCCS}
S_{\rm {NCCS}}={1\over 4\pi\nu}\ \int_M \epsilon^{\mu\nu\alpha} {\rm
Tr}\left(A_{\mu}\star\parnu A_{\alpha} +{2i\over 3} A_{\mu}\star
A_{\nu}\star A_{\alpha}\right)\ ,
\ee
where $\star$-product is defined in (\ref{starp}), $A_{\mu}$ take
values in $U(N)$ algebra (unitary $N\times N$ matrices) and ${\rm Tr}$ is
basically trace over the $U(N)$ indices. Here we consider the base
manifold $M$ which can 
be separated into a two dimensional (\nc) subspace times $\rr$ or $S^1$
and further we assume that we can define the $\star$-product introduced
earlier
(\ref{starp}), i.e. we consider the \nc plane, \nc cylinder
\cite{{NCcyl},{Chaich}}
and \nc torus. However, we should remind ourselves that for the case
of torus, depending on the fact that the \ncy per unit volume is rational or
irrational, one can realize two completely different cases. 
For the rational \nc torus, our gauge theory (\nc \CS) can be mapped into   
a commutative theory with a non-trivial background flux while
it is not possible for the irrational \nc torus \cite{NCT}.
Here we do not consider the torus case in detail and will only make some
remarks passing at the end of section 3 and 4.

The action (\ref{NCCS}) is invariant under infinitesimal gauge
transformations:
\be\label{smallG}
A_{\mu}\longrightarrow A'_{\mu}=A_{\mu}+\parmu\lambda+i(\lambda\star
A_{\mu}- A_{\mu}\star\lambda)\ ,\ \ \ \ \ \lambda\in U(N)\ .
\ee
However, the case of finite gauge transformations should be studied
separately and this we
will do in the next section.

Analogous to the \com case, one can define the \nc Yang-Mills-\CS theory
by
adding the \nc Yang-Mills action to (\ref{NCCS}) \cite{Wu}.
The perturbative loop calculations have also been done  for both pure NCCS
\cite{Bichl} and for \nc Yang-Mills-\CS \cite{Wu}.  The pure \CS
action at one loop level can once again be recovered by sending the gauge
coupling to infinity.

The key observation in performing the \nc loop calculations in general, is
that the $\star$-product (and hence the Moyal bracket) in the Fourier
modes take a simple form, namely
$$
\{{\rm e}^{ik\cdot x}, {\rm e}^{ip\cdot x}\}=2i\sin({1\over 2}k\theta p)     
{\rm e}^{i(k+p)\cdot x}\ ,\ \ \ \ k\theta
p=k_{\mu}\theta^{\mu\nu}p_{\nu}\ , 
$$
and therefore the loop integrals in general can be decomposed in two
parts, one which contains the $\theta$ factor (${\rm e}^{ik\theta p}$),
the non-planar part, and one which do not depend on $\theta$, the planar
part
\cite{{MRS},{Tomb}}. At one loop level the non-planar parts are (UV)
finite while they show an IR divergence. This is known as IR/UV mixing
which is a peculiar feature of \nc field theories \cite{{MRS},{Tomb}}.

For the gauge theories however, the divergent part of the loop integrals,
compared to the \com counter-part, contains an extra factor of 
$4\sin^2{({1\over 2}k\theta p)}=2(1+\cos(k \theta p))$. The loop
integrals lead to the usual (\com) $\beta$-function 
equation but the $C_2(G)$ (quadratic Casimir of the gauge group) is now
replaced by $2\times C_2(G)$ \cite{Ren} which for NC$U(N)$ that is $2N$.

For the \nc \CS theory ({\it as a limit of} \nc Yang-Mills-\CS ) the loop 
calculations have been performed and shown that \cite{Wu}
\footnote{Actually the results of \cite{Wu} differ with ours  in the
factor of 2. That is related to the {\it improper normalization} they use.}
\be\label{NCren} 
{1\over \nu_{{\rm ren}}}={1\over \nu_{{\rm class}}}+ 2\times N\ .
\ee
The novel point shown explicitly in \cite{Wu} is that all the non-planar
diagrams vanish in the pure \CS limit, i.e. we are {\it not} 
going to find any further IR divergence due to these terms. However, if we
just start with the pure \CS we will persist with the IR divergences
coming from non-planar diagrams \cite{Bichl}.
This can be understood by noting that the \CS term introduces a
topological mass term in
the Yang-Mills action \cite{Jackiw}, and this mass term serves as a natural IR
regulator, which, even in the pure \CS limit, tames the IR divergences of
non-planar diagrams. Presumably the
eq.(\ref{NCren}), although  being a one loop result, holds for all loop
orders, similar to the commutative case \cite{Das}. In fact this is
expected
because we can again recognize the BRST and vector supersymmetry in this
case \cite{{Bichl},{D-Sh}}. 

The eq.(\ref{NCren}) is also remarkable noting the statistics of the \nc
\CS particles \cite{Susskind}: the quantum corrections to ${1\over \nu}$
are not
going to change the statistics of these particles, this is in fact what is
expected from a well-defined quantum theory.

The other important property of the action (\ref{NCCS}) is that it is independent of
the choice of metric on $M$ (note that metric is not involved in $\star$-product).
However, in order to be treated as a topological field theory one should guarantee
that this property remains at the level of the quantum partition function. On the
other hand to perform the path integral we should fix the gauge. Adding the gauge
fixing and ghost terms to the action spoils the metric independence, 
and hence the topological nature of the theory becomes questionable
\cite{B-T1}. In the usual
\com \CS theories this is overcome recalling the BRST symmetry of the full
gauge fixed action and the fact 
that the gauge fixing and the ghost part of the action can be expressed as
commutator of BRST charge with a
local function of the
fields \cite{B-T1}.
This argument also holds for the \nc case. The existence of the BRST symmetry have
been shown in \cite{Bichl}. Also there it has been shown that the metric dependent
(gauge fixing) part  of the action is a BRST {exact} form
\footnote{Besides the action, in the path integral there is also the
measure, which is the same  for a \com theory and its \nc counter-part
\cite{Andrei}. However, as has been discussed in \cite{B-T2} the measure
is metric independent.}. This guarantees the metric
independence of the \nc \CS at quantum level. Hence, the \nc\CS can be treated as a
topological field theory. 
\newsection{The invariance  of \nc \CS  \\[-0.6cm] 
under finite gauge transformations}

In this section we would like to address another feature of \nc \CS
theories on the $\rr^2_{\theta}\times \rr$. 
As we have discussed the \nc \CS action (\ref{NCCS}) is invariant under
small gauge
transformations. However, it is also important to study its behaviour
under finite \nc gauge transformations defined by
\be\label{finite}
A_{\mu}\longrightarrow A^g_{\mu}=g^{-1}\star
A_{\mu}\star g +ig^{-1}\star \parmu g\ ,
\ee
where $g$ is valued in NC$U(N)$ group: $g$ is an $N\times N$ matrix satisfying
\be\label{gg}
g^{\dagger}\star g= g\star g^{\dagger}= {\bf 1}\ .
\ee
One can show that 
\ba\label{lambda}
g&=&({\rm e}\star)^{i\lambda}\equiv 1+i\lambda-{1\over 2}\lambda\star\lambda-{i\over
3!}\lambda\star\lambda\star\lambda+ \cdots ,\cr
g^{\dagger}&=& g^{-1}=({\rm e}\star)^{-i\lambda}\ , 
\ea
with $\lambda$ being a $N\times N$ matrix, solves (\ref{gg}).
We recall that, according to definition of $NCU(N)$ algebras  $g^{-1}dg$
should be a member of the algebra of the
compact operators on $\rr^2_{\theta}\times \rr$ \cite{Rieffel}.  
In addition, we also impose the boundary condition 
\be\label{compact}
g_{_{x\to\infty}}\to{\bf 1}.
\ee

Under the above gauge transformations the action (\ref{NCCS}) transforms as
$S\to S+\delta_g S$, where
\be
4\pi\nu\ \delta_g S=-\int \epsilon^{\mu\nu\alpha}\Bigl[\parmu {\rm Tr}(\parnu g\star
g^{-1}\star A_{\alpha})-{i\over 3} 
{\rm Tr}(g^{-1}\star\parmu g\star g^{-1}\star\parnu g \star
g^{-1}\star\partial_{\alpha} g)\Bigr] .
\ee
The first term in the above is a total derivative and vanishes due to
boundary conditions, while the second term
is a non-trivial one, usually called the "winding number" of the group. For
the usual
compact groups, this winding number is an integer. However, for the \nc case one
should redo the corresponding computations, and this is what we do in the
following.  In order to show how the method works, let us first consider the
NC$U(1)$ case, and then we will generalize it to NC$U(N)$.
\vskip .3cm
{\it The NC$U(1)$ winding number}
\vskip .5cm

In this case $g$ is just a function (without matrix indices), but we should 
remember that in general it is $\theta$ dependent.
First taking derivatives of (\ref{gg}) we find two very useful identities:
\be\label{gmug} 
\parmu g^{-1}\star g =- g^{-1}\star \parmu g\ ,
\ee
and 
\be\label{gthetag} 
{-i\over 4}\partial_{[\mu}g^{-1}\star \partial_{\nu]} g
={\delta\over\delta{\theta^{\mu\nu}}}g^{-1}\star g +
g^{-1} \star {\delta\over\delta{\theta^{\mu\nu}}}g\ .
\ee
The identity (\ref{gthetag}) is obtained by taking the derivative of
(\ref{gg}) with
respect to $\theta_{\mu\nu}$. Using the identity (\ref{gmug})  we find
$$
g^{-1}\star\parmu g\star g^{-1}\star\parnu g=-\parmu g^{-1}\star \parnu g\ .
$$ 
Then we use the identity (\ref{gthetag}), and again
(\ref{gmug}). Recalling the boundary condition (\ref{compact}) we can 
drop the total derivative term and we remain with
$$
\int \epsilon^{\mu\nu\alpha}\ (g^{-1}\star\parmu g\star g^{-1}\star\parnu g
\star g^{-1}\star\partial_{\alpha} g)= 4i\int \epsilon^{\mu\nu\alpha}\
{\delta\over\delta{\theta^{\mu\nu}}}\left(g^{-1}\star\partial_{\alpha}
g\right)\ .
$$
If we parameterize the $\rr$ part by $t$ and the $\rr^2_{\theta}$ by $x,y$,  we have
\be\label{deltaU(1)}
\delta_g S={-i\over 2\pi\nu}\epsilon^{ij}{\delta\over\delta\theta^{ij}} \int
dxdy \int dt \left(g^{-1} \partial_t g\right)\ ,
\ee
where $i,j=1,2$. 
Now, noting the representation of $g$, (\ref{lambda}), one can prove that
\be\label{U(1)}
g^{-1}\star\partial_{\alpha} g=i\partial_{\alpha}\lambda +\{ B_{\alpha},
\lambda\}\ ,
\ee
where $B_{\alpha}$ is a function which essentially is made out of
$\lambda$ and 
$\partial_{\alpha}\lambda$. Inserting the above in (\ref{deltaU(1)}), the
second term in (\ref{U(1)}) which is a Moyal bracket vanishes upon
integrating over $x,y$. Then the integration over $t$ can be easily
performed and we find
\be\label{deltaU1}
\left. \delta_g S={1\over 2\pi\nu}\epsilon^{ij}{\delta\over\delta\theta^{ij}}
\int dxdy\ \lambda\right|_{t=-\infty}^{t=+\infty}\ .
\ee
To evaluate the $x,y$ integration it is more convenient to use the
operatorial language developed in \cite{Harvey}:
$$
\int dx dy \longleftrightarrow \pi\epsilon_{ij}\theta^{ij}\ ``{\rm tr}" ,
$$ 
where $``{\rm tr}"$ is the trace over the Hilbert space of corresponding
harmonic oscillator basis and any function $\lambda$ can be expanded in terms
of harmonic oscillator states
$$
\lambda(t)=\sum a_{mn}(t)\mid n\rangle\langle m\mid \ .
$$
The boundary conditions (\ref{compact}) is satisfied if
\be
a_{mn}(t\to \infty)=\delta_{m,n}\ 2\pi K_n\ ,\ \ \ \ K_n\in \zz \ . 
\ee
Now, we have all the ingredients, inserting these all in (\ref{deltaU(1)})
\ba
\delta_g S&=&{1\over 2\pi\nu}\epsilon^{ij}\ \pi\epsilon_{ij}\ 2\pi K \cr 
          &=& {2\pi K\over\nu}\ ,\ \ \ \ \ K\in \zz\ .
\ea
However, in order to have a well-defined quantum theory $\delta_g S$ should
be an integer multiple of $2\pi$, i.e. 
\be
{1\over \nu}\in \zz\ .
\ee
So, we see that in the \nc case, the level of the \CS should be quantized. 

\vskip .3cm
{\it The NC$U(N)$ winding number}
\vskip .5cm

All the above steps, before (\ref{U(1)}) work all the same for the NC$U(N)$ case.
However, the (\ref{U(1)}) can be replaced with another identity:
\be\label{U(N)}
{\rm Tr}(g^{-1}\star\partial_{\alpha} g)=i{\rm Tr} \partial_{\alpha}\lambda
+\{
B^a_{\alpha}, \lambda^a\}\ ,
\ee
where $^a$ stands for the $U(N)$ group indices. Hence, again upon
plugging back (\ref{U(N)}) in the (\ref{deltaU(1)}), we face the level 
quantization. 

So, we have shown that the winding number for the \nc plane case is non-zero
and leads to level quantization. This result have also been observed using another
calculational method \cite{Nair}.
We note that our result and also our arguments heavily rely on non-zero
$\theta$, furthermore the limit $\theta\to 0$ is not a smooth one. In
particular for the NC$U(1)$ case again we need the level quantization.

It would be very interesting if we can generalize the general result
of \com case, namely $\Pi_3(G)=\zz$ for any compact group $G$, to the 
\nc case in more mathematical language \cite{prog}. 

For the \nc torus case of course, we should note that, since again there 
are some gauge transformations that are not smoothly connected to ${\bf 1}$, 
and as it has been discussed in
\cite{Thomas}, the winding number will still be an integer and
hence in the torus case again we require the level quantization condition.
We will discuss the torus case later in the next section.

\newsection{Discussions and remarks}

In this work we have studied some aspects of \nc \CS theories. First we
discussed and reviewed the perturbative loop calculations. We discussed
that we do not face the IR/UV mixing (the non-planar diagrams vanish), if
we start with the \nc Yang-Mills-\CS and then after performing the loop
calculations
send the gauge coupling to infinity to obtain the pure \nc \CS
action. We should recall that this limit is well-defined because (similar to 
the \com case) we expect the quantum corrections on the \CS coupling
${1\over \nu}$ to appear only at one loop level, and the one loop result
(\ref{NCren}) remain to all higher order loops.   
However, starting from the pure \CS we still find the non-planar IR
divergences \cite{Bichl}. The (\ref{NCren}) is remarkable remembering 
the statistics of the \nc \CS particles
(\ref{anyon}) \cite{Susskind}.
Altogether it seems that for the NC$U(1)$ case, values of
${1\over \nu}$ differing by an even integer are physically
equivalent.
Studying the behaviour of the \nc\CS under finite gauge transformations
we have proved its invariance up to some integer multiple of 
the \CS coupling, and hence in order the \nc \CS theory to make sense 
at quantum level the coupling (level of \CS theory) should be quantized.
We would like to stress that our proof and
hence our result for NC$U(1)$ case does not hold in the $\theta^{\mu\nu}=0$
limit, i.e. the $\theta\to 0$ is not a smooth limit.

Here we have mainly discussed the \CS model on \nc plane.
For the rational \nc torus case because
there is a finite dimensional representation for the coordinates and
generators of translations \cite{{Chaich},{Szabo}}, namely for
$\Theta={P\over Q}$ we have
a $Q\times Q$ representation, the  arguments should be revised.
However, we can still make some statements about the rational \nc torus case.
Using the Morita equivalence, one can map $NCU(1)$ \CS into a
$U(Q)$ \CS theory on a usual \com torus  but with a non-zero magnetic
flux on the torus \cite{NCT}, i.e. we should add some terms linear in the
gauge field to the \CS action.  In other words, the classical
equations of motion will not be the flat connections anymore. Instead we
have to deal with the connections of constant curvature. The moduli space of
these connections have also been studied well and we expect that it is possible to
work out the precise form of the partition function for this case. We postpone this
to future work \cite{prog}.

In the \nc case there is a possible extension for the \CS theory.
In general since we have another two form in our problem, namely
$\theta^{-1}_{\mu\nu}=\omega_{\mu\nu}$, we can add the term like
$$
\int {\rm
Tr}\left(\epsilon^{\mu\nu\alpha}\omega_{\mu\nu}A_{\alpha}\right)\
$$
to the \CS action, where $\omega_{\mu\nu}$ defines a constant 2-form which
in a sense measures the \ncy of the space.  
We note that in 3 dimensions $\theta^{\mu\nu}$ has a zero  eigenvalue, and so by
$\theta^{-1}_{\mu\nu}$, we mean the inverse of $\theta$ 
in the \nc $2\times 2$ block. So the extended action reads as
\be
S={1\over 4\pi\nu}\ \int \epsilon^{\mu\nu\alpha} {\rm
Tr}\left(A_{\mu}\star\parnu A_{\alpha} +{2i\over 3} A_{\mu}\star
A_{\nu}\star A_{\alpha}\right)\ + {i\gamma\over 4\pi} \int {\rm
Tr}\left(\epsilon^{\mu\nu\alpha}\omega_{\mu\nu}A_{\alpha}\right)\ ,
\ee 
where $\gamma$ is a dimensionless parameter, which is  related to the unit of
monopole charge. We also note that the gauge invariance of this action now
implies that ${1\over\nu}+\gamma\in \zz$. The possibility of having
this additional term (which has
no analog in the \com case) has the significant effect of changing the
equations of motion for the connection from the zero curvature to a
constant curvature (proportional to $\omega$). 

We also discussed that the \nc \CS theory can be treated as a topological
field theory. Then the next step to explore these theories is to find
their partition function, and also the Wilson loops. 
It would also be interesting to consider the 
theory on the other \nc manifolds, such as \nc sphere$\times\rr$ or
$\Sigma\times\rr$, with
$\Sigma$ being a \nc surface of arbitrary genus.
However, our discussions in this work 
depend crucially on the definition of the $\star$-product, (\ref{starp}), and the
fact that $\theta$ is a constant. 

If one can work out the partition function for \nc \CS theories, besides
the mathematical interests, this can be very interesting physically in the
context of quantum Hall effect \cite{Susskind}.

\vskip 1cm

{\bf Acknowledgements}

I would like to thank A. Mukherjee for long discussions and
comments and also for his collaboration at early stages of this work. 
I am grateful to R. Hernandez, M. Schnabl, M. Blau, George Thompson and  
especially to L. Susskind and T. Krajewski for fruitful discussions and
comments.

\end{document}